\documentclass[a4paper,fleqn,usenatbib]{mnras}

\usepackage[T1]{fontenc}
\usepackage{ae,aecompl}

\usepackage{graphicx}	
\usepackage{amsmath}	
\usepackage{amssymb}	
\usepackage{ulem}       
\usepackage{hyperref}
\usepackage{subcaption}

\title[Sombrero galaxy stream]{A feather on the hat: Tracing the giant stellar stream around the Sombrero galaxy}

\author[D.\ Mart{\'\i}nez-Delgado et al.]{David
Mart{\'\i}nez-Delgado$^{1}$\thanks{Talentia Senior Fellow}\thanks{E-mail: dmartinez@iaa.es},
Javier Román$^{1,2,3}$, Denis Erkal $^{4}$, Mischa Schirmer$^{5}$,\newauthor Santi Roca-F\`abrega$^{6}$, Seppo Laine$^{7}$, Giuseppe Donatiello$^{8}$, Manuel Jimenez$^{9}$,\newauthor  David Malin$^{10}$, Julio A. Carballo-Bello$^{11}$  \\\\
$^{1}$ Instituto de Astrof\'isica de Andaluc\'ia, CSIC, E-18080, Granada, Spain\\
$^{2}$ Instituto de Astrof\'{\i}sica de Canarias, c/ V\'{\i}a L\'actea s/n, E-38205, La Laguna, Tenerife, Spain\\
$^{3}$ Departamento de Astrof\'{\i}sica, Universidad de La Laguna, E-38206, La Laguna, Tenerife, Spain\\
$^{4}$Department of Physics, University of Surrey, Guildford GU2, 7XH, UK \\
$^{5}$ Max-Planck Institute for Astronomy, K\"onigstuhl 17, 69117, Heidelberg, Germany\\
$^{6}$Departamento de F{\'\i}sica de la Tierra y Astrof{\'\i}sica, Universidad Complutense de Madrid, E-28040 Madrid, Spain\\
$^{7}$ IPAC, Mail Code 314-6, Caltech, 1200 E. California Blvd., Pasadena, CA 91125 USA\\
$^{8}$UAI - Unione Astrofili Italiani /P.I. Sezione Nazionale di Ricerca Profondo Cielo, 72024 Oria, Italy \\
$^{9}$ MJ Observatory, Cuenca, Spain\\
$^{10}$ Former AAO, Macquarie University, 105 Delhi Rd, North Ryde, NSW 2113, Australia\\
$^{11}$ Instituto de Alta Investigaci\'on, Sede Esmeralda, Universidad de Tarapac\'a, Av. Luis Emilio Recabarren 2477, Iquique, Chile\\
}

\date{Accepted XXX. Received YYY; in original form ZZZ}

\pubyear{2018}

\begin{document}
\label{firstpage}
\pagerange{\pageref{firstpage}--\pageref{lastpage}}
\maketitle

\begin{abstract}
Recent evidence of extremely metal-rich stars found in the Sombrero galaxy (M104) halo suggests that this galaxy has undergone a recent major merger with a relatively massive galaxy. In this paper,  we present wide-field deep images of the M104 outskirts obtained with a 18-cm amateur telescope with the purpose of detecting any coherent tidal features from this possible major merger. Our new data, together with a model of the M104 inner halo and scattered light from stars around the field, allow us to trace for the first time the full path of the stream on both sides of the disk of the galaxy. We fully characterize the ring-like tidal structure and we confirm that this is the only observable coherent substructure in the inner halo region. This result is in agreement with the hypothesis that M104 was created by a wet major merger more than 3.5\,Gyr ago that heated up the stellar population, blurring all old substructure. We generated a set of numerical models that reproduce the formation of the observed tidal structure. Our best fit model suggests the formation of this stream in the last 3\,Gyr is independent of the wet major merger that created the M104 system. Therefore, the formation of the tidal stream can put a constraint on the time when the major merger occurred. 

\end{abstract}

\begin{keywords}
galaxies:individual: M104 -- galaxies: interaction -- galaxies: formation -- galaxies: haloes
\end{keywords}


\section{Introduction}

Recent deep, wide-area photometric surveys have revealed a large number
of faint stellar substructures (``streams'') around spiral galaxies,
resulting from the tidal disruption of lower-mass galaxies. 
While detailed studies of resolved streams around the MW and M31 imply a
dynamic hierarchical accretion history, consistent with $\Lambda$CDM cosmological galaxy
formation models
\citep[e.g.,][]{bullock05,delucia08,cooper10,cooper13,pillepich15,rodriguez16}, a much larger sample of galaxies 
is needed to test whether the merging histories of the Local
Group spirals are typical of galaxies in their mass range \citep[e.g.,][]{mutch11,morales18}.

A crucial ingredient in testing whether the merger histories of the MW and M31 are
typical (consistent with $\Lambda$CDM cosmological galaxy formation models) is the
acquisition of  adequately deep images, as the majority of the predicted tidal stellar
streams have surface brightnesses in the R-band fainter than about 29 AB mag arcsec$^{-2}$.
While a few deep imaging surveys of the outskirts of local galaxies have recently been
completed \citep[e.g.,][]{tal09,delgado10,ludwig12,duc15} or are ongoing \citep{2020ApJ...894..119D, 2021arXiv210406071M},
the majority of nearby galaxies have not been observed down to the surface brightnesses needed
to detect streams from ancient minor mergers. 

\begin{figure*}
	\includegraphics[width=1.0\textwidth]{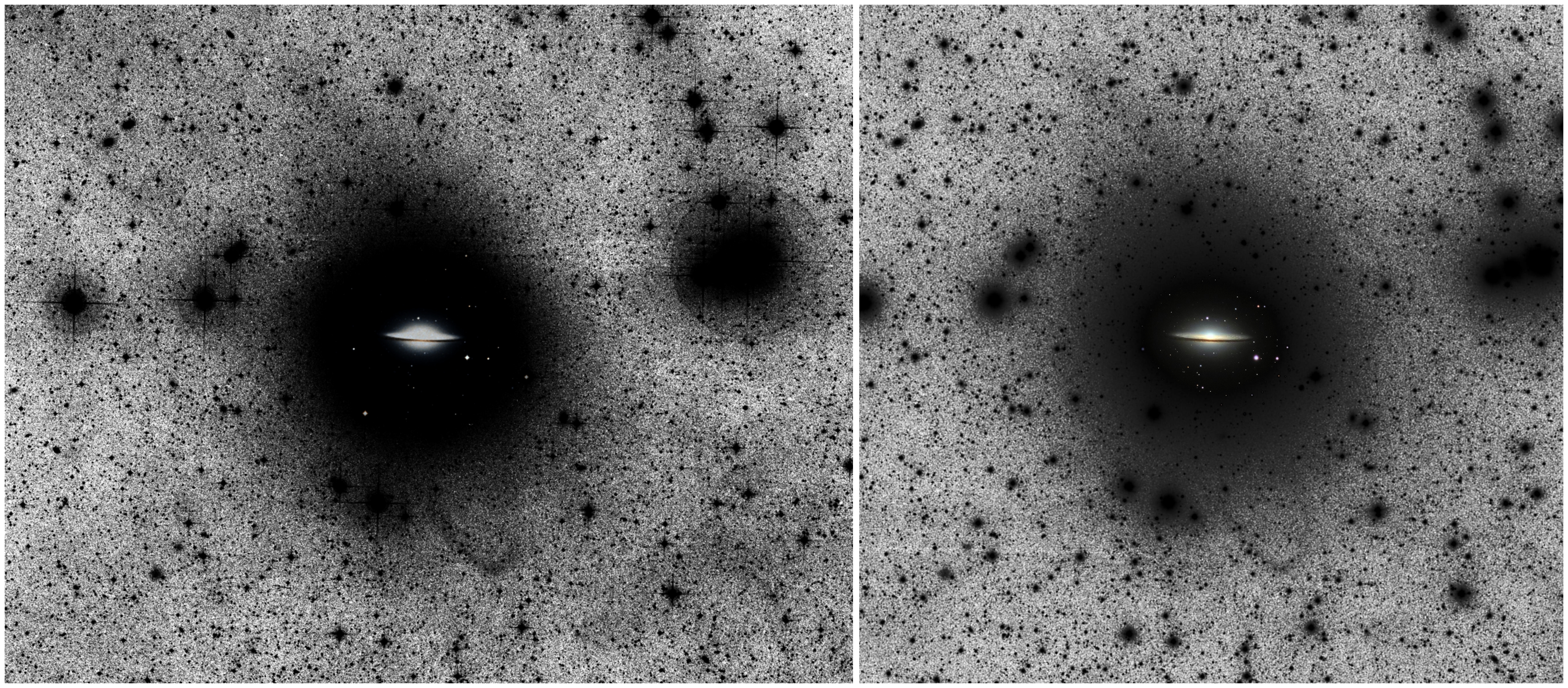}
    \caption{{\it Left panel:} Original discovery photographic plate of M104 stream by \citep{1997PASA...14...52M}. {\it Right panel:} Luminance-filter image of the M104 stream obtained with an 18-cm apochromatic refractor telescope and a total exposure time of 4.75 hours. The total field-of-view of this image is 50\arcmin $\times$ 50\arcmin. North is up and East is left.}
    \label{fig-malin-ccd}
\end{figure*}

The halos in the simulations of \citet{bullock05} typically have about two streams brighter
than 30 AB mag arcsec$^{-2}$\,\footnote{All magnitudes are in the AB system \citep{oke_1983}}. However, the majority of substructures are at surface
brightnesses fainter than 30 AB mag arcsec$^{-2}$. Our current inability to see the fainter
streams (corresponding to either earlier merger epochs or lower mass progenitors) implies
that currently the merger history that we study in galaxies beyond the Local Group is
strongly biased towards the most recent (the last few tens of percent of mass accretion)
and/or the most massive minor merger events. Thus we are only sensitive to the most metal-rich populations, as
validated by studies of resolved stars around M31 \citep{mcconnachie18} and Cen A
\citep{crnojevic16}. Close to the center of a galaxy (R$_{\rm proj}$ $<$ 30 kpc) the
substructure is most likely generated by the most massive merged satellite galaxies, as
dynamical friction will have brought them quickly to the central regions. Therefore, our
current view of tidal streams in nearby galaxies is highly biased towards the most massive
minor mergers which are relatively rare for MW-like galaxies.


Stellar tidal streams around all but the most nearby galaxies cannot be resolved into stars with our modest telescopes, and thus appear as elongated diffuse light regions that extend over several arc minutes as projected  on the sky.  Their typical surface brightness is 26 mag arcsec$^{-2}$ or fainter, depending on the luminosity of the progenitor and the time they were accreted \citep{johnston01}. Detecting these faint features requires very dark sky conditions and wide-field, deep images taken with exquisite flat-field quality over a wide region ($>$ 30\arcmin) around the targets. By focusing on nearby spiral galaxies with diffuse-light over-densities, the Stellar Tidal Stream Survey (STSS) has discovered more than 50 previously unknown stellar structures so far at distances $<$ 50 Mpc \citep{delgado18}. The morphologies of these diffuse-light structures  include ``great circle'' streams, isolated shells, giant debris clouds,
jet-like features, and large diffuse structures that may be old, phase mixed remnants of merged companions. Again, very similar features
are seen in cosmological simulations of minor mergers \citep{johnston08,cooper10}.

One of the first cases of stellar streams reported beyond the Local Group is M104 (NGC 4594, also known as the ``Sombrero galaxy"). This massive galaxy is located in a low density environment in the direction of the Virgo cluster. Recent estimates of its total mass have been obtained by using radial velocities and projected separations of some of its satellite dwarf galaxies. These estimates suggest M104 has a total mass of 1.55$\pm$0.49$\times$10$^{13}$M$_{\odot}$ \citep[][]{Karachentsev2020}, a value that is in good agreement with previous studies \citep[e.g. see mass estimates using orbit-based models in][]{Jardel2011}. A distance of $\sim$ 9.5 Mpc has been computed using the tip of the red giant branch method  (TRGB) in several  of its confirmed satellite galaxies \citep{m104_distance,Karachentsev2020}.
Its peculiar morphology has puzzled researchers for many years, resembling a combination of a massive halo, that includes a massive bulge/inner halo plus a more extended spheroid/external halo, and a dusty disk nearly edge-on. It was initially classified as an Sa, a nearly edge-on, galaxy \citep{deVaucouleurs1991, Emsellem1996}, but recently re-classified as an elliptical galaxy by \citet{Gadotti2012}. \citet{Gadotti2012} made a detailed structural analysis of the M104 disk, bulge and halo components by using Spitzer IRAC 3.6$\mu$m imaging, and concluded that this galaxy is an outlier in scaling relations of disk galaxies when all components are considered together. However, when considered independently, both the disk and the bulge-halo components fit well with regular disk and intermediate-mass elliptical galaxies, respectively \citep{Cohen2020}.
The disk component shows a low and very low star formation rate in the inner disk and dust ring, respectively. This result suggests an ongoing inside-out star formation quenching process \citep{DeLooze2012}.

More intriguing is the halo--bulge system. Most observations show that its globular cluster population follows the bi-modal distribution in the color/metallicity space observed in several other spiral and elliptical galaxies \citep[e.g][]{Harris2017}, but more interesting is that the two GC populations also show clear differences in their kinematics \citep{Jardel2011}. These differences in the kinematics clearly point to a different origin. Additionally, although the blue GC population shows a metal distribution that coincides with the one observed in giant elliptical galaxies, the red population is much more metal-rich than expected for an elliptical galaxy, and does not follow the expected mass--metallicity relation \citep{Harris2010,Harris2015,Cohen2020}. On the other hand, a growing number of satellite galaxy systems are being discovered in the galaxy neighbourhood. So far, 15 companions have been confirmed, one of them being an Ultra Compact Dwarf that is deeply embedded in the M104 inner halo \citep{Hau2009}. Most of these recently discovered satellite galaxies are low surface brightness systems \citep[e.g.][]{Javanmardi2016,Karachentsev2020} but still need to be confirmed by precise distance determinations \citep{Carlsten2020}.

Several formation scenarios have been proposed in order to explain the unusual morphology and properties of M104, both of its elliptical and disk components. \citet{Gadotti2012} proposed a two-stage formation with a former gas inflow that generated the old ellipsoid and a more recent inflow of pre-enriched gas that built up the disk. Although this scenario successfully explains the formation of the observed bi-modal GC distribution, it is not supported by the current knowledge of cosmic gas accretion models in which galaxies as massive as M104 do not allow new cold gas inflows from the cosmic web \citep{Birnboim2003}.  

More recently,  \citep{Cohen2020} proposed that the extremely metal-rich stars observed in the M104 halo can be produced in a recent major merger with a log(M/M$_{\odot}$)$\sim$11.1 galaxy. In this scenario, the metal-poor blue GC population was part of an old elliptical galaxy while the metal-rich red GCs, and most of the metal-rich halo stars, were accreted from a companion.  Motivated by this recent study, we have obtained new deep, wide-field images of M104 to look for traces of cohererent tidal features or shells which could confirm this major merger scenario.


\section{Observations and Data Reduction}

\subsection{Discovery photographic plate}

The first deep image of M104  revealing its stellar stream and some faint extensions was made by combining photographically-amplified data from two UK Schmidt Telescope (UKST) plates taken in blue light \citep{1997PASA...14...52M}. A traditional silver halide photographic plate was coated with a thin layer of gelatin containing a three-dimensional dispersion of tiny, light-sensitive microcrystals, distributed uniformly throughout the `emulsion'. When exposed to light, then processed in a liquid developer, the light-struck crystals are reduced to metallic silver grains, their number (and thus the optical density) is proportional to the relative intensity of the incoming radiation. The silver halide crystals strongly scatter visible light within the emulsion, such that upper layers of the emulsion always contain the faintest images. In contact-copying plates with a diffuse light to make a positive, high contrast photographic image, this property can be exploited to emphasize the faintest details on a plate (``photographic amplification"), at the expense of brighter features. The process is non-destructive, and was found to be especially useful for revealing very faint details in photographic optical astronomy \citep{1978Natur.276..591M, 1979Natur.277..279M}. A full technical description of this photographic amplification is given in \cite{1981AASPB..27....4M} . 

The left panel of Fig.~\ref{fig-malin-ccd} shows the resulting photographic plate of the M104 stream. The well-defined South-West loop is clearly visible and there is evidence of a fainter, large, but less well defined extension to the North-East. Later, three further red-light images were added, which confirmed its existence. On all these images, including the new one here, the well-defined southern loop appears detached from the inner diffuse halo of the galaxy. These features are about $\sim$ 5.5 magnitudes fainter than the natural dark sky and extend about 100 kpc on either side of the galaxy. 
This stellar stream was subsequently confirmed with imaging from the Sloan Digital Sky Survey \citep{2011A&A...536A..66M}.

\subsection{Wide-field imaging with a small telescope}

The STSS has established a successful search strategy to detect
tidal streams to very faint surface brightness limits. This survey strategy aims to obtain multiple deep
exposures of the target galaxy with modest-sized (0.1-0.5-meter) telescopes using a latest
generation astronomical commercial CCD camera. The images are obtained using high throughput clear 
filters with near-IR cut-off, known as luminance filters \citep[e.g., see Fig. 1 in][]{delgado15} 
and long exposures of at least 7--8 hours. 

The right panel of Fig.~\ref{fig-malin-ccd} shows an image of a wide field around M104 obtained during 2019 and 2020 with a
 CFF 180mm F/7 apochromatic refractor with Astro-Physics QUADTCC reducer (yielding an effective ratio of F/5.2) remotely operated in Cuenca (Spain). It used a Moravian G3 16200 CCD camera and Chroma Luminance filter with a pixel scale of 1.32".

The reduction was carried out with the usual steps of bias and dark subtraction. The flat-fielding was done in Pixinsight, using an Artesky 550-mm flat-field generator. Astrometry was obtained using \texttt{SCAMP} \citep{2006ASPC..351..112B}. A final stacked image was obtained by combining 57$\times$300-s best images, with a total exposure time of 285 minutes (4.75 hours). The photometric calibration was referenced to the \textit{r} band of the Sloan Digital Sky Survey (SDSS) in this field using a mosaic from the SDSS DR12 mosaic utility. The surface brightness limit of the image is $\mu_{lim,r}$ = 27.3 mag arcsec$^{-2}$ measured as 3$\sigma$ in 10$\times$10 arcsecond boxes \citep[following the recipe by][appendix A]{2020A&A...644A..42R}, which is approximately one magnitude deeper than the SDSS DR8 images.

\subsection{Observations with the 4.2m William Herschel Telescope}
On 2009 April 22, we observed the stellar stream in M104 with the Prime Focus Imaging Camera at the 4.2 m William Herschel Telescope (WHT) on La Palma, Spain. The images were taken in the Harris R filter. The first hours of observations were impacted by severe high clouds resulting in up to 0.8 mag extinction; these exposures were rejected to minimize spurious signals from scattering in the background. The remaining images were taken in photometric conditions with a total integration time of $12\times600 = 7200$ s, with a seeing of $0.92^{\prime\prime}$ in the final coadded image.

\begin{figure}
	\includegraphics[width=1.0\columnwidth]{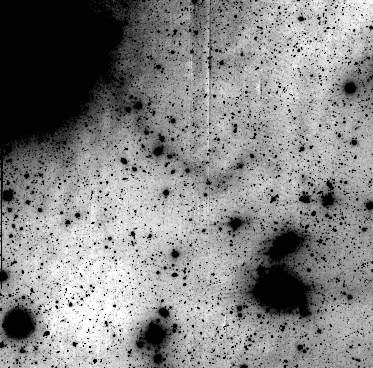}
    \caption{WHT 4.2-m image of the M104 stream. North is up and East is left, the field of view is $15^\prime\times15^\prime$.}
    \label{fig-wht}
\end{figure}

The data were reduced using the pipeline THELI v3.0.4 \citep{sch13}, using standard bias and flat-fielding pre-processing to remove the instrumental fingerprint. The astrometric calibration and distortion correction for the individual exposures was performed against Gaia DR2 \citep{gaia_mission,gaia_dr2}. The image series was interleaved with offset exposures to a blank sky field nearby, from which we planned to extract a background correction and fringe map. However, when processing the data, it became evident that the dither pattern for the blank sky field was not wide enough to account for the presence of brighter stars in the area. This led to numerous local over-estimations of the background on angular scales comparable to those of the stream, despite the advanced masking techniques available in THELI. We therefore decided to not use the data from the blank sky fields.
The fringing amplitude in the data is on the order of 0.5\%, on a spatial scale $2-10\times$ smaller than the width of the stream and of comparable brightness. Coadding the dithered exposures averaged out the fringing signal to a negligible level. The coadded image featured a residual approximately linear gradient of 2\% of the background value in the vertical direction, which we removed using column averaging. The stream is well visible in the data (Fig. \ref{fig-wht}). 

\begin{figure}
	\includegraphics[width=1.0\columnwidth]{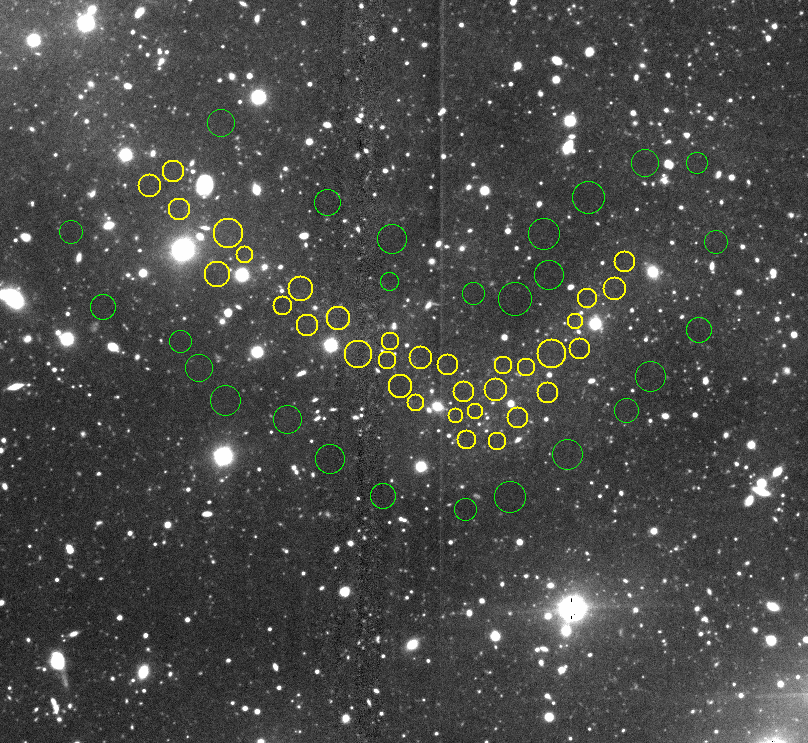}
    \caption{Placement of apertures to determine the mean local background (green circles) and the mean stream surface brightness (yellow circles) of the apocenter of the stream. The field of view corresponds to the inner half of Fig. \ref{fig-wht}}.
    \label{fig-stream_surface_brightness}
\end{figure}

Flux calibration of the coadded image was achieved by cross-matching unsaturated stars against  PanSTARRS-DR1 $g$ and $r$-band photometry \citep{panstarrs_dr1}, using a linear fit for the color term. We set the diameter of the photometric apertures to 50 pixels ($11^{\prime\prime}.8$), where the growth curve leveled off and contamination by neighboring sources was still negligible. The photometric calibration equation for the coadded image in the AB mag system is 
    {\rm ZP} = 27.18 + 0.0454\,(g-r),
with a statistical uncertainty of $\sigma({\rm ZP})=0.039$ AB mag.

\section{Tracing the stellar stream in the inner halo of M~104}
\label{sec:stream}

\begin{figure*}
	\includegraphics[width=1.0\textwidth]{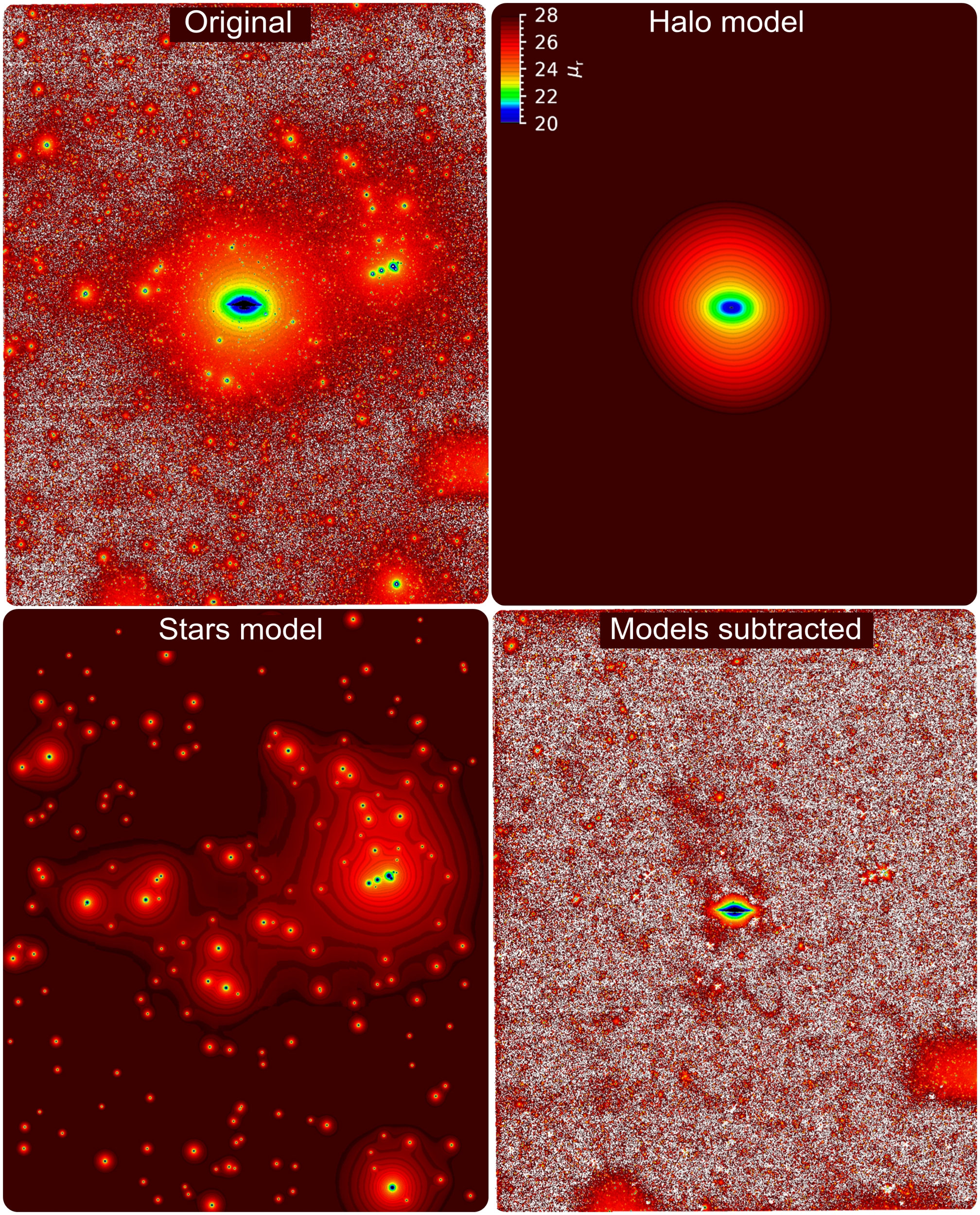}
    \caption{Modelling and subtraction processes. {\it Upper left panel}: Coadded image of M104 obtained in this work. {\it Upper right panel}: Halo model. {\it Bottom left panel}: Stars model. {\it Bottom right panel}: Original image subtracted from models. Images have been enhanced by a gaussian smoothing of 5 pixels kernel. The surface brightness scale is common to all panels and is shown in the upper right panel. Some residual light due to ghost light from nearby bright stars remains on the edges of the image, having not been modelled.}
    \label{fig:model}
\end{figure*}

In Fig. \ref{fig:model}, left panel, we show the final coadded image from the small telescope (see Sec. 2.2). It is evident that there is a considerable confusion due to the scattered light from the stars present in the field. Additionally, the prominent halo of M104 obstructs any analysis of features that may be located in the area adjacent or over the galaxy, including the low surface brightness tidal stream that appears visible in the outermost area of the halo at south-west direction. In order to expose any low surface brightness features buried beneath the confusing light from the M104 halo or stars present in the field, we carried out a modeling of the sources of confusion. 

First, we built a model of the scattered light from nearby stars in the field. For this, we carried out specific observations of bright stars to produce a high signal-to-noise, large-radius PSF model. We follow a processing similar to that carried out by \citet{2020A&A...644A..42R,2020MNRAS.491.5317I}. Given the excellent circular symmetry of the PSF of this telescope, we performed azimuthal symmetry of the model in order to maximize the signal-to-noise, which produces our final PSF model. Finally we run the scattered-light modelling pipeline developed by \citet{2020A&A...644A..42R} producing the model of the light scattered by the stars in the field (see Fig. \ref{fig:model}). The number of fitted stars is 176. This is all the stars in the field up to magnitude 13.5 in the \textit{r} band and is more than enough to eliminate confusion by bright stars in the field.

For the modelling of the stellar halo of M104 we performed a masking of all external sources. Additionally, and after exhaustive preliminary tests, we masked the region of the central dust disk and its surroundings. The reason is 
because we find it difficult to make a fit to this morphology due to its complexity, so we prefer to model only the outer region or halo of M104. For the modeling we use \texttt{IMFIT} \citep{2015ApJ...799..226E} on the masked image. We obtained a very satisfactory result in modeling the halo by using two Sersic functions, one of them with a `boxy' parameter. The first function manages to fit the innermost part and closest to the disc, with parameters: PA = 88.1$^{\circ}$, n = 1.11, ellip = 0.42, c$_{0}$ = 0.38 and R$_{eff}$ = 135.6\arcsec using a boxy Sersic function. The second function fits to the outermost region of the halo with parameters of PA = 17.9$^{\circ}$, n = 1.06 and R$_{eff}$ = 337.5\arcsec. The total magnitude of the halo model (both functions, see Fig. \ref{fig:model}) is 8.18 mag. 

The result of the subtraction of the models of the light scattered by stars and the M104 halo can be seen in the lower right panel of Fig. \ref{fig:model}. The modeling produces very clean residuals, leaving only the innermost region of the M104 disc with S0 type morphology and a clearly visible diffuse feature looping around M104. Due to the regions of oversubtraction near the S0 disk of M104, the integrated photometry of the whole stream is infeasible. We have calculated the maximum surface brightness of the stream, in the southern region, resulting in approximately $\mu _{r}$ = 26.7 $\pm$ 0.4 mag arcsec$^{-2}$. An estimate of the color of the feature is not possible in our data in which we only have the luminance filter. We have explored other data sets, for example in the Canada--France--Hawaii Telescope archive, with which to obtain a color in this brighter region, however it is infeasible due to its low surface brightness and the strong background fluctuations that make the photometry unsuitable for a robust color determination.

The surface brightness of the possible apocenter of the stream was determined by manually placing circular apertures on the stream (situated at 18.70\arcmin ~from the center of M104) and in its local vicinity (for background determination), see also Fig. \ref{fig-stream_surface_brightness}. We find a mean surface brightness of 
\begin{eqnarray}
    \mu_R & = & (4.16\pm0.04\pm0.15)\times10^{-2}\;\mu{\rm Jy\;arcsec^{-2}}\\
    & = & 27.351\pm0.010\pm0.039\;{\rm AB\;mag\;arcsec}^{-2}\;.
\end{eqnarray}
The first error is the measurement uncertainty, including the fluctuations in the background samples and the stream samples shown in Fig. \ref{fig-stream_surface_brightness}. The second error is systematic, propagated from the uncertainty of the photometric zeropoint.


\section{Stream simulations} \label{sec:simulations}

\begin{figure*}
  \includegraphics[width=1.0 \textwidth]{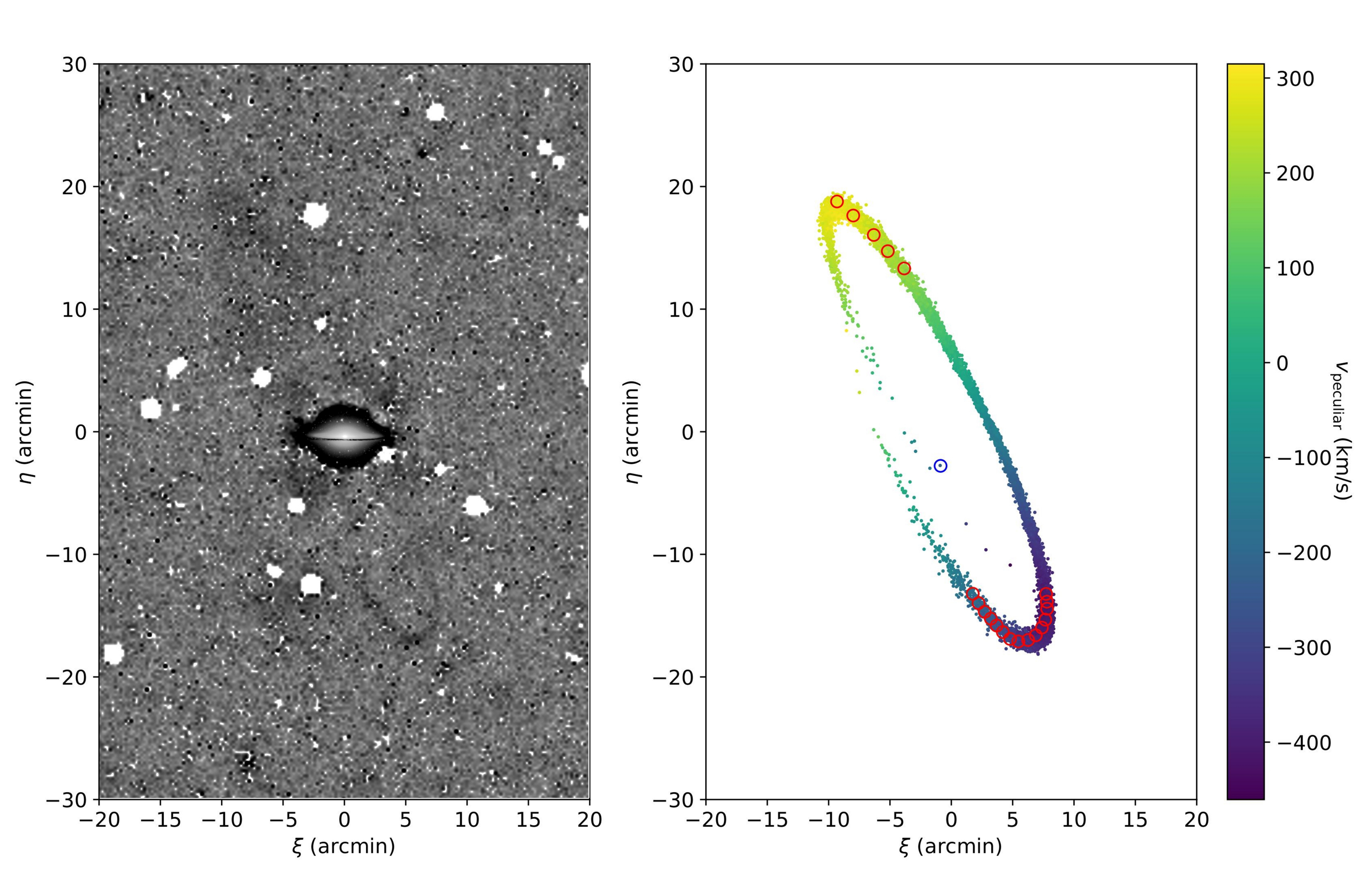}
\caption{Comparison of the M104 stream with a stream model. \textit{Left panel} shows the observed stream around M104. This image has been rebinned to enhance the signal-to-noise of the stream. \textit{Right panel} shows the best-fit stream model colored by the predicted peculiar velocity. The red circles shows the observed stream track locations and the black points show the simulated stream. The blue circle shows the location of the recently discovered ultra-compact dwarf. This best-fit model shows that it is possible to reproduce the observed tidal features around M104 with a single disruption event.}
\label{fig:stream_fit}
\end{figure*}

In this section, we fit a stream model to the M104 stream to assess whether the lower and upper stream could plausibly be a single structure. We first measure the stream track, using the image showed in the left panel of Figure \ref{fig:stream_fit}, which has been re-binned to enhance its signal-to-noise. Using this figure, the stream track and width are measured by eye at several locations along the lower and upper stream. The locations of the track measurements are depicted by the red circles in the right panel of Figure \ref{fig:stream_fit}.

In order to generate stream models, we use the modified Lagrange Cloud stripping (mLCS) technique of \cite{gibbons14}. This technique allows for the rapid generation of realistic streams which can then be fit to the observations. For the potential, we use a potential motivated by the results of \cite{Jardel2011}. In particular, we use an exponential disk with a mass of $6.9\times10^{10} M_\odot$, a scale radius of $1.9$ kpc, and a scale height of $0.3$ kpc, a Hernquist bulge \citep{hernquist1990} with a mass of $1.91\times10^{11} M_\odot$ and a scale radius of $9$ kpc, and a cored logarithmic halo with a circular velocity of $374$ km/s and core radius of $4.7$ kpc. This potential closely matches the enclosed mass measurement of \cite{Jardel2011} beyond a radius of $\sim 20$ kpc. We model the progenitor as a Plummer sphere \citep{plummer1911} with an initial mass of $10^8 M_\odot$ and a scale radius of $500$ pc. We note that these progenitor parameters were chosen to match the observed stream width. 

We parametrize our stream model in terms of the progenitor's present-day position and velocity relative to M104. These Cartesian coordinates are aligned with our view of M104, i.e., the $+x$ direction points to the West, the $+y$ direction points to the North, and the $+z$ direction points towards the Sun with an origin at the center of M104. For simplicity, we place the progenitor at $y=0$ (i.e., close to the plane of the disk), leaving five parameters for the progenitor's present day coordinates. Once the stream is generated in these coordinates, we convert into mock observations (i.e., angles on the sky) by placing the center of our simulated M104 at a distance of 9.8 Mpc \citep[in agreement with][]{Jardel2011}. Note that we also account for a disk inclination of $80^\circ$ relative to the line of sight \citep[in agreement with][]{Jardel2011} by inclining the disk in the simulation.

Given the progenitor's present-day position, we rewind the progenitor's orbit for 4 Gyr and then disrupt the stream along this orbit using the mLCS technique. We compute the likelihood of each stream model by comparing it with the observed track. This approach is motivated by fits to streams in the Milky Way \citep[e.g.][]{erkal19}. We do this by computing the stream's on-sky coordinates, and convert these into polar coordinates centered on M104, ($r,\theta$), where $r$ is the on-sky distance from M104 and $\theta$ is the polar angle measured counter-clockwise from the West direction. For each measurement of the observed stream, we use a Gaussian likelihood of

\begin{equation}
\log \mathcal{L} = - \sum_i \Big( \log \sqrt{2\pi \sigma^2} + \frac{(r_i^{\rm obs}-r_i)^2}{2 \sigma^2} \Big),
\end{equation}
where $r_i^{\rm obs}$ is the observed radius of the stream track, $r_i$ is the mean radius of the simulated stream within one degree (measured in the on-sky polar coordinates, $\theta$) of the i$^{\rm th}$ track location, and $\sigma$ is an additional nuisance parameter in our model to account for the uncertainty of the stream track locations. Thus our model has a total of six parameters.

In order to explore the likelihood surface, we use the Markov Chain Monte Carlo (MCMC) approach as implemented in \textsc{emcee} \citep{emcee}. We use uniform priors on all of our parameters with the only requirement being that $\sigma > 0$. Before starting the MCMC, we find a set of parameters which roughly match the stream by trial and error. We initialize the walkers close to these values. For the MCMC, we use 100 walkers, 1000 steps, and a burn-in of 500 steps. We stress that the goal of this fitting procedure is only to show that there is a physically motivated stream model consistent with the observations of M104 stream. We do not claim that this is the only such model and we note that without any observations of the velocity, there are multiple, discrete solutions allowed. For example, for any model, the velocities can be flipped which will reverse the direction of the stream and give nearly the same stream locations of the sky. We also note that all of the parameters were constrained by the MCMC and there were no unconstrained degeneracies (i.e., no parameters or combination of parameters which the likelihood does not depend on). We confirmed with a longer MCMC run (100 walkers, 10000 steps, 5000 step burn in) where we fit an orbit to the observed stream.

We show the best-fit model in the right panel of Figure \ref{fig:stream_fit}. The stream progenitor is on an eccentric orbit with a pericenter of $\sim36$ kpc, an apocenter of $\sim65$ kpc, and an orbital period of $\sim 600$ Myr. The orbit of this stream is inclined by $\sim77^\circ$ relative to the line of sight so the stream is seen close to edge-on. We also note that while the stream model shown here was integrated for 4 Gyr, if this integration is reduced to 3 Gyr, the stream is still long enough to match the observed stream extent. Thus, we have shown that the stream features seen around M104 are consistent with a single disruption event. Future radial velocity measurements would be helpful to confirm this picture and we note that since the stream is close to edge-on, in parts of the stream the predicted radial velocities relative to M104 are substantial ($\sim 400$ km/s).

In Figure \ref{fig:stream_fit}, we also denote the location of the massive ultra-compact dwarf (UCD)  reported by \citet{Hau2009} with a blue circle. Interestingly, the dwarf's location is close to the leading arm of the stream. However, if this UCD is the progenitor of the stream, there should be a (roughly) equally long leading and trailing stream emanating from the dwarf. Given that the dwarf sits on the edge of the best-fit model stream, and thus that the dwarf would have an asymmetric stream, we thus conclude that this compact object is likely not connected to the stream. 



\section{Discussion}

In this paper, we present new deep images of the Sombrero galaxy (M104) outskirts obtained with a 18-cm telescope,  with the aim of detecting tidal features from a possible major merger recently proposed by \citet{Cohen2020}. Our new data allows us to trace the full path
of the stream on both sides of the disk of the galaxy, suggesting it is the remnant of a tidally disrupted dwarf galaxy accreted in on a modestly eccentric orbit roughly 3 Gyr ago. Unfortunately, the lack of accurate broad-band photometry prevents us from measuring its total luminosity or to gain any insights into its stellar population. The path of the stream does not intersect with any of the two halo fields  recently targeted by the HST \citep{Cohen2020} to probe the stellar metallicity gradient of the M104 stellar halo. Thus, there is no contamination from this stellar stream that could explain the strikingly metal-rich population ([Z/H]$\sim$-0.15) of this halo reported in that study. Apart from this well-known, `great-circle'-type stellar stream emerging outside the halo,  our analysis does not show any additional tidal feature embedded in its bright halo. 

In this context it is important to know if this tidal feature was produced by the same processes that created the M104's unusual disk and halo structures, or if it was the result of a more recent minor merger. According to recent results by \citet{Cohen2020} the M104 system was a result of a wet merger, i.e., the companion was a gas-rich disky galaxy, taking place $>$3.5\,Gyr ago \citep[][]{McNeilMoylan2012,ElicheMoral2018,Mancillas2019}. In this hypothesis the incoming galaxy brought to the resulting system enough gas and dust to rejuvenate the old elliptical (the main progenitor), and to rebuild a gaseous/dusty disk. After this merger two globular cluster populations, one from each progenitor, were left in the halo, a bi-modal distribution that is observed in the M104 system \citep{Tonini2013}. This major merger scenario is also in a good agreement with observations of the AGN activation process and the following disk’s star formation quenching and final AGN activity decline, that is, this final step, in where M104 currently is \citep{Catalan2017,Sanchez2018}. The major merger scenario proposed here has also been considered to explain properties of other massive galactic systems in the local universe \citep[e.g.][]{McNeilMoylan2012}. For example, this scenario can explain the short transit from early-type to late-type-like system observed in galaxies showing exotic morphology, like the NGC 3108 \citep[see][]{Hau2008}. Assuming this major merger scenario is correct, the resulting system would show small stellar substructure in its inner halo and disk region. Cosmological simulations and theoretical models have shown that strong interactions speed up the phase mixing that ends up diluting all transient structures in galactic halos of kinematically hot elliptical galaxies, i.e. tidal tails, stellar streams and shells \citep{Feldmann2008,Bellstedt2017}. This picture is in a good agreement with the lack of substructure we observed. 

It is well known that major mergers can also create ring-like structures when the interacting galaxies collide with a particular impact parameter \citep{Freeman1974}. However, we reject the possibility that the observed tidal structure is a collisional ring generated during the major merger. This was because, (a) the homogeneity of its stellar, gas, globular clusters, and planetary nebulae distributions suggest that the galactic system is already relaxed \citep{Ford1996,Harris2010} and (b), because the orientation, distance to the central galaxy, and thickness of the stream are not compatible with a major merger origin \citep{Wang2012,ElicheMoral2018,2021arXiv210406071M}. 

In the light of our results and of the predictions from theoretical models, we suggest that this stream is not a result of the processes that created the M104’s peculiar disk structure and halo but an independent minor merger event. A direct consequence of this hypothesis is that the formation of the stream would set a lower limit for the time when the M104 was assembled. According to \citet[][]{Mancillas2019} this happened between 3.5 and 3.0\,Gyr ago, which is in good agreement with our best-fit stream model in Section~\ref{sec:simulations}.

Our best-fit stream model in Section \ref{sec:simulations} can match the extent of the observed stream with a disruption time of 3\,Gyr. The stream's location in the dense inner halo region also suggests that the progenitor of this transient structure could be a compact dwarf system that succeeded in penetrating the outer halo and reached this dense region before being stripped out \citep[see e.g.][]{BoylanKolchin2007,Oogi2013}. An interesting candidate is the UCD identified by \citet{Hau2009} in HST/ACS imaging, and confirmed to be associated with the Sombrero galaxy by its recession velocity obtained from Keck spectra. However, the connection between the detected stellar stream and the UCD cannot be further explored without kinematical data of the stream, something that is challenging, even with a 8-meter class telescope due to its extremely low surface-brightness even at its apocenter ($\mu_R$ = 27.35; see Sec.~\ref{sec:stream}). One possibility to improve our stream simulation is to identify the known globular clusters of M104 \citep{2007ApJ...658..980B} likely associated to the stream with available radial velocities, following a similar approach to that used in the NGC 5907 stream by \citet{2020MNRAS.491.5693A}. We will address this issue in a forthcoming study of the M104 stream.

Finally, it is important to mention that the gas-rich major merger scenario needs to be confirmed in future observations of M104's luminosity and kinematic profiles. Many previous studies have proposed that the differences in the assembly history of S0 galaxies drive the characteristic luminosity and kinematic profiles \citep{Cox2006,Borlaff2014,Tapia2014}. Of special interest will be to study the projected ellipticities, rotation parameter, kinematic misalignment, and spin parameter of the three differentiated components of M104, i.e., the outer halo, the inner halo, and the stellar disk \citep{Cox2006}. An accurate characterization of its physical properties, and a detailed comparison of those with results from theoretical models and simulations will help to better understand the origin of this peculiar galactic system.

\section*{Acknowledgements}
DMD devotes this paper to Prof. Jose Luis Comellas, who guided him to find the Sombrero galaxy from his home rooftop in the 1980s. 
DMD acknowledges financial support from the Talentia Senior Program (through the incentive ASE-136) from Secretar\'\i a General de  Universidades, Investigaci\'{o}n y Tecnolog\'\i a, de la Junta de Andaluc\'\i a. DMD and JR acknowledge funding from the State Agency for Research of the Spanish MCIU through the {\it Center of Excellence Severo Ochoa} award to the Instituto de Astrof{\'i}sica de Andaluc{\'i}a (SEV-2017-0709). 
JR acknowledges support from the State Research Agency (AEI-MCINN) of the Spanish Ministry of Science and Innovation under the grant "The structure and evolution of galaxies and their central regions" with reference PID2019-105602GB-I00/10.13039/501100011033
DMD also recognizes funds from the research project AYA 2007-65090. THELI made use of the following tools and data products: Source Extractor \citep{bertin96}, Scamp \citep{2006ASPC..351..112B}, Swarp \citep{2002ASPC..281..228B}, and the VizieR catalogue access tool, CDS, Strasbourg, France. Based on observations made with the William Herschel Telescope operated on the island of La Palma by the Isaac Newton Group of Telescopes in the Spanish Observatorio del Roque de los Muchachos of the Instituto de Astrofísica de Canarias.

This work has made use of data from the European Space Agency (ESA) mission
{\it Gaia} (\url{https://www.cosmos.esa.int/gaia}), processed by the {\it Gaia}
Data Processing and Analysis Consortium (DPAC,
\url{https://www.cosmos.esa.int/web/gaia/dpac/consortium}). Funding for the DPAC
has been provided by national institutions, in particular the institutions
participating in the {\it Gaia} Multilateral Agreement.

\section*{Data Availability}

The data underlying this article will be shared on a reasonable request to the corresponding author.



\bibliographystyle{mnras}

\begin{thebibliography}{99}
\bibitem[Alabi et al.(2020)]{2020MNRAS.491.5693A} Alabi, A.~B., Forbes, D.~A., Romanowsky, A.~J., et al.\ 2020, \mnras, 491, 5693. doi:10.1093/mnras/stz3382
\bibitem[Amorisco(2015)]{amorisco15} Amorisco, N. C. 2015, MNRAS, 450, 575
\bibitem[\protect\citeauthoryear{Bellstedt et al.}{2017}]{Bellstedt2017} Bellstedt S., Forbes D.~A., Foster C., Romanowsky A.~J., Brodie J.~P., Pastorello N., Alabi A., et al., 2017, MNRAS, 467, 4540. doi:10.1093/mnras/stx418
\bibitem[Bertin \& Arnouts(1996)]{bertin96} Bertin, E. \& Arnouts, S.\ 1996, \aaps, 117, 393. doi:10.1051/aas:1996164
\bibitem[Bertin et al.(2002)]{2002ASPC..281..228B} Bertin, E., Mellier, Y., Radovich, M., et al.\ 2002, Astronomical Data Analysis Software and Systems XI, 281, 228
\bibitem[Bertin(2006)]{2006ASPC..351..112B} Bertin, E.\ 2006, Astronomical Data Analysis Software and Systems XV, 351, 112
\bibitem[Birnboim \& Dekel(2003)]{Birnboim2003} Birnboim, Y. \& Dekel, A.\ 2003, \mnras, 345, 349. doi:10.1046/j.1365-8711.2003.06955.x
\bibitem[\protect\citeauthoryear{Borlaff et al.}{2014}]{Borlaff2014} Borlaff A., Eliche-Moral M.~C., Rodr{\'\i}guez-P{\'e}rez C., Querejeta M., Tapia T., P{\'e}rez-Gonz{\'a}lez P.~G., Zamorano J., et al., 2014, A\&A, 570, A103. doi:10.1051/0004-6361/201424299
\bibitem[\protect\citeauthoryear{Boylan-Kolchin \& Ma}{2007}]{BoylanKolchin2007} Boylan-Kolchin M., Ma C.-P., 2007, MNRAS, 374, 1227. doi:10.1111/j.1365-2966.2006.11276.x
\bibitem[Bridges et al.(2007)]{2007ApJ...658..980B} Bridges, T.~J., Rhode, K.~L., Zepf, S.~E., et al.\ 2007, \apj, 658, 980. doi:10.1086/511413
\bibitem[Bullock \& Johnston(2005)]{bullock05} Bullock, J. S. \& Johnston, K. V. 2005,
ApJ, 635, 931
\bibitem[Carlsten et al.(2020)]{Carlsten2020} Carlsten, S.~G., Greco, J.~P., Beaton, R.~L., et al.\ 2020, \apj, 891, 144. doi:10.3847/1538-4357/ab7758
\bibitem[Catal{\'a}n-Torrecilla et al.(2017)]{Catalan2017} Catal{\'a}n-Torrecilla, C., Gil de Paz, A., Castillo-Morales, A., et al.\ 2017, \apj, 848, 87. doi:10.3847/1538-4357/aa8a6d
\bibitem[Chambers et al.(2016)]{panstarrs_dr1} Chambers, K.~C., Magnier, E.~A., Metcalfe, N., et al.\ 2016, arXiv:1612.05560
\bibitem[Cohen et al.(2020)]{Cohen2020} Cohen, R.~E., Goudfrooij, P., Correnti, M., et al.\ 2020, \apj, 890, 52. doi:10.3847/1538-4357/ab64e9
\bibitem[Cooper et al.(2010)]{cooper10} Cooper, A. P., et al. 2010, MNRAS, 406, 744
\bibitem[Cooper et al.(2013)]{cooper13} Cooper, A. P., et al. 2013, MNRAS, 434, 3348
\bibitem[\protect\citeauthoryear{Cox et al.}{2006}]{Cox2006} Cox T.~J., Dutta S.~N., Di Matteo T., Hernquist L., Hopkins P.~F., Robertson B., Springel V., 2006, ApJ, 650, 791. doi:10.1086/507474
\bibitem[Crnojevi\'{c} et al.(2016)]{crnojevic16} Crnojevi\'{c}, D., et al. 2016, ApJ, 823, 19
\bibitem[Danieli et al.(2020)]{2020ApJ...894..119D} Danieli, S., Lokhorst, D., Zhang, J., et al.\ 2020, \apj, 894, 119. doi:10.3847/1538-4357/ab88a8
\bibitem[De Looze et al.(2012)]{DeLooze2012} De Looze, I., Baes, M., Fritz, J., et al.\ 2012, \mnras, 419, 895. doi:10.1111/j.1365-2966.2011.19759.x
\bibitem[De Lucia \& Helmi(2008)]{delucia08} De Lucia, G. \& Helmi, A. 2008, MNRAS, 391, 14
\bibitem[de Vaucouleurs et al.(1991)]{deVaucouleurs1991} de Vaucouleurs, G., de Vaucouleurs, A., Corwin, H.~G., et al.\ 1991, Third Reference Catalogue of Bright Galaxies. Volume I: Explanations and references.  Volume II: Data for galaxies between 0$^{h}$ and 12$^{h}$. Volume III: Data for galaxies between 12$^{h}$ and 24$^{h}$., by de Vaucouleurs, G.; de Vaucouleurs, A.; Corwin, H. G., Jr.; Buta, R. J.; Paturel, G.; Fouqu{\'e}, P.. Springer, New York, NY (USA), 1991, 2091 p., ISBN 0-387-97552-7, Price US\$ 198.00. ISBN 3-540-97552-7, Price DM 448.00. ISBN 0-387-97549-7 (Vol. I), ISBN 0-387-97550-0 (Vol. II), ISBN 0-387-97551-9 (Vol. III).
\bibitem[Duc et al.(2015)]{duc15} Duc, P.--A., et al. 2015, MNRAS, 446, 120
\bibitem[\protect\citeauthoryear{Eliche-Moral et al.}{2018}]{ElicheMoral2018} Eliche-Moral M.~C., Rodr{\'\i}guez-P{\'e}rez C., Borlaff A., Querejeta M., Tapia T., 2018, A\&A, 617, A113. doi:10.1051/0004-6361/201832911

\bibitem[Emsellem et al.(1996)]{Emsellem1996} Emsellem, E., Bacon, R., Monnet, G., et al.\ 1996, \aap, 312, 777
\bibitem[Erkal, Sanders \& Belokurov(2016)]{erkal16} Erkal, D., Sanders, J. L., Belokurov, V. 2016, MNRAS, 461, 1590
\bibitem[Erkal et al.(2019)]{erkal19} Erkal, D., Belokurov, V., Laporte, C.~F.~P., et al.\ 2019, \mnras, 487, 2685. doi:10.1093/mnras/stz1371
\bibitem[Erwin(2015)]{2015ApJ...799..226E} Erwin, P.\ 2015, \apj, 799, 226
\bibitem[Feldmann et al.(2008)]{Feldmann2008} Feldmann, R., Mayer, L., \& Carollo, C.~M.\ 2008, \apj, 684, 1062. doi:10.1086/590235
\bibitem[\protect\citeauthoryear{Ford et al.}{1996}]{Ford1996} Ford H.~C., Hui X., Ciardullo R., Jacoby G.~H., Freeman K.~C., 1996, ApJ, 458, 455. doi:10.1086/176828
\bibitem[Foreman-Mackey et al.(2013)]{emcee} Foreman-Mackey, D., Hogg, D.~W., Lang, D., et al.\ 2013, \pasp, 125, 306. doi:10.1086/670067
\bibitem[\protect\citeauthoryear{Freeman \& de Vaucouleurs}{1974}]{Freeman1974} Freeman K.~C., de Vaucouleurs G., 1974, ApJ, 194, 569. doi:10.1086/153276
\bibitem[Gadotti \& S{\'a}nchez-Janssen(2012)]{Gadotti2012} Gadotti, D.~A. \& S{\'a}nchez-Janssen, R.\ 2012, \mnras, 423, 877. doi:10.1111/j.1365-2966.2012.20925.x
\bibitem[Gaia Collaboration et al.(2016)]{gaia_mission} Gaia Collaboration, Prusti, T., de Bruijne, J.~H.~J., et al.\ 2016, \aap, 595, A1. doi:10.1051/0004-6361/201629272
\bibitem[Gaia Collaboration et al.(2018)]{gaia_dr2} Gaia Collaboration, Brown, A.~G.~A., Vallenari, A., et al.\ 2018, \aap, 616, A1. doi:10.1051/0004-6361/201833051
\bibitem[Gibbons, Belokurov, \& Evans(2014)]{gibbons14} Gibbons, S. L. J., Belokurov,
\bibitem[Grand et al.(2016)]{grand17} Grand, R.J.J. et al. 2017, MNRAS, 467, 179
\bibitem[Hellwing et al.(2016)]{hellwing16} Hellwing, W. A., et al. 2016, MNRAS, 457, 3492
\bibitem[Harris et al.(2017)]{Harris2017} Harris, W.~E., Ciccone, S.~M., Eadie, G.~M., et al.\ 2017, \apj, 835, 101. doi:10.3847/1538-4357/835/1/101
\bibitem[Harris et al.(2015)]{Harris2015} Harris, W.~E., Harris, G.~L., \& Hudson, M.~J.\ 2015, \apj, 806, 36. doi:10.1088/0004-637X/806/1/36
\bibitem[Harris et al.(2010)]{Harris2010} Harris, W.~E., Spitler, L.~R., Forbes, D.~A., et al.\ 2010, \mnras, 401, 1965. doi:10.1111/j.1365-2966.2009.15783.x
\bibitem[Hau et al.(2009)]{Hau2009} Hau, G.~K.~T., Spitler, L.~R., Forbes, D.~A., et al.\ 2009, \mnras, 394, L97. doi:10.1111/j.1745-3933.2009.00618.x
\bibitem[Hau et al.(2008)]{Hau2008} Hau, G.~K.~T., Bower, R.~G., Kilborn, V., et al.\ 2008, \mnras, 385, 1965. doi:10.1111/j.1365-2966.2007.12740.x
\bibitem[Hendel \& Johnston(2015)]{hendel15} Hendel, D. \& Johnston, K. V. 2015, MNRAS, 454, 2472
\bibitem[Hendel et al.(2018)]{hendel18} Hendel, D., et al. 2018, arXiV:1811:10613
\bibitem[Hernquist(1990)]{hernquist1990} Hernquist, L.\ 1990, \apj, 356, 359. doi:10.1086/168845
\bibitem[Infante-Sainz et al.(2020)]{2020MNRAS.491.5317I} Infante-Sainz, R., Trujillo, I., \& Rom{\'a}n, J.\ 2020, \mnras, 491, 5317
\bibitem[Jardel et al.(2011)]{Jardel2011} Jardel, J.~R., Gebhardt, K., Shen, J., et al.\ 2011, \apj, 739, 21. doi:10.1088/0004-637X/739/1/21
\bibitem[Javanmardi et al.(2016)]{Javanmardi2016} Javanmardi, B., Martinez-Delgado, D., Kroupa, P., et al.\ 2016, \aap, 588, A89. doi:10.1051/0004-6361/201527745
\bibitem[Johnston et al.(2001)]{johnston01} Johnston, K. V., Sackett, P. D., Bullock, J. S. 2001, ApJ, 557, 137
\bibitem[Johnston et al.(2008)]{johnston08} Johnston, K. V., et al. 2008, ApJ, 689, 936
\bibitem[Kado--Fong et al.(2018)]{kado-fong18} Kado--Fong, E., et al. 2018, ApJ, 866, 103
\bibitem[Karachentsev et al.(2020)]{Karachentsev2020} Karachentsev, I.~D., Riepe, P., \& Zilch, T.\ 2020, Astrophysics, 63, 5. doi:10.1007/s10511-020-09608-5
\bibitem[Karachentsev et al.(2020)]{Karachentsev2020} Karachentsev, I.~D., Makarova, L.~N., Tully, R.~B., et al.\ 2020, arXiv:2008.13152
\bibitem[Karachentsev et al.(2009)]{karachentsev09} Karachentsev, I. D., Kashibadze, O. G., Makarov, D. I., \& Tully, R. B. 2009, MNRAS, 393, 1265
\bibitem[Laine et al.(2016)]{laine16} Laine, S., et al. 2016, AJ, 152, 72
\bibitem[Ludwig et al.(2012)]{ludwig12} Ludwig, J., Pasquali, A., Grebel, E. K., \& Gallager, J. S. III 2012, AJ, 144, 190
\bibitem[Malin(1978)]{1978Natur.276..591M} Malin, D.~F.\ 1978, \nat, 276, 591. doi:10.1038/276591a0
\bibitem[Malin(1979)]{1979Natur.277..279M} Malin, D.~F.\ 1979, \nat, 277, 279. doi:10.1038/277279a0
\bibitem[Malin(1981)]{1981AASPB..27....4M} Malin, D.~F.\ 1981, AAS Photo Bulletin, 27, 4
\bibitem[Malin \& Hadley(1997)]{1997PASA...14...52M} Malin, D. \& Hadley, B.\ 1997, \pasa, 14, 52. doi:10.1071/AS97052
\bibitem[Mancillas et al.(2019)]{Mancillas2019} Mancillas, B., Duc, P.-A., Combes, F., et al.\ 2019, \aap, 632, A122. doi:10.1051/0004-6361/201936320
\bibitem[Mart\'{i}nez Delgado et al.(2010)]{delgado10} Mart\'{i}nez--Delgado, D.,
et al. 2010, AJ, 140, 962
\bibitem[Mart\'{i}nez Delgado et al.(2012)]{delgado12} Mart\'{i}nez--Delgado, D.,
et al. 2012, ApJ, 748, 24
\bibitem[Mart\'{i}nez Delgado et al.(2015)]{delgado15} Mart\'{i}nez--Delgado, D.,
et al. 2015, AJ, 150, 116
\bibitem[Mart\'{i}nez Delgado (2018)]{delgado18} Mart\'{i}nez Delgado, D. 2018, in {\it Highlights of the Spanish Astrophysics X}, Montesinos, B., Asensio-Ramos, A., Buitrago, F., Schodel, R., Villaver, E., Peréz-Hoyos, S. (eds.), in press
\bibitem[Martinez-Delgado et al.(2021)]{2021arXiv210406071M} Martinez-Delgado, D., Cooper, A.~P., Roman, J., et al.\ 2021, arXiv:2104.06071
\bibitem[McConnachie et al.(2018)]{mcconnachie18} McConnachie, A. W., et al. 2018, ApJ, in press (arXiV:1810:08234)
\bibitem[McQuinn et al.(2016)]{m104_distance} McQuinn, K.~B.~W., Skillman, E.~D., Dolphin, A.~E., et al.\ 2016, \aj, 152, 144. doi:10.3847/0004-6256/152/5/144
\bibitem[\protect\citeauthoryear{McNeil-Moylan et al.}{2012}]{McNeilMoylan2012} McNeil-Moylan E.~K., Freeman K.~C., Arnaboldi M., Gerhard O.~E., 2012, A\&A, 539, A11. doi:10.1051/0004-6361/201117875
\bibitem[Miskolczi et al.(2011)]{2011A&A...536A..66M} Miskolczi, A., Bomans, D.~J., \& Dettmar, R.-J.\ 2011, \aap, 536, A66. doi:10.1051/0004-6361/201116716

\bibitem[Mihos et al.(2017)]{mihos17} Mihos, J. C., et al. 2018, ApJ, 834, 16
\bibitem[Morales et al.(2018)]{morales18} Morales, G., et al. 2018, A\&A, in press (arXiV:1804:03330)
\bibitem[Mutch, Croton, \& Poole(2011)]{mutch11} Mutch, S. J., Croton, D. J., \& Poole, G. B. 2011, ApJ, 736, 84
\bibitem[Oke \& Gunn(1983)]{oke_1983} Oke, J.~B. \& Gunn, J.~E.\ 1983, \apj, 266, 713
\bibitem[\protect\citeauthoryear{Oogi \& Habe}{2013}]{Oogi2013} Oogi T., Habe A., 2013, MNRAS, 428, 641. doi:10.1093/mnras/sts047
\bibitem[Pillepich, Madau, \& Mayer(2015)]{pillepich15} Pillepich, A., Madau, P, \& Mayer, L. 2015, ApJ, 799, 184
\bibitem[Pillepich et al.(2018)]{pillepich18} Pillepich, A., et al. 2018, MNRAS, 473, 4077
\bibitem[Plummer(1911)]{plummer1911} Plummer, H.~C.\ 1911, \mnras, 71, 460. doi:10.1093/mnras/71.5.460
\bibitem[Rodr\'{i}guez--G\'{o}mez et al.(2016)]{rodriguez16} Rodr\'{i}guez--G\'{o}mez, V., et al. 2016, MNRAS, 458, 2371
\bibitem[Rom{\'a}n et al.(2020)]{2020A&A...644A..42R} Rom{\'a}n, J., Trujillo, I., \& Montes, M.\ 2020, \aap, 644, A42. doi:10.1051/0004-6361/201936111

\bibitem[S{\'a}nchez et al.(2018)]{Sanchez2018} S{\'a}nchez, S.~F., Avila-Reese, V., Hernandez-Toledo, H., et al.\ 2018, \rmxaa, 54, 217
\bibitem[Sanders \& Binney(2013)]{sanders13} Sanders, J. L., \& Binney, J. 2013, MNRAS, 433, 1826
\bibitem[Schirmer(2013)]{sch13} Schirmer, M.\ 2013, \apjs, 209, 21. doi:10.1088/0067-0049/209/2/21
\bibitem[Slater et al.(2009)]{slater09} Slater, C. T., Harding, P., Mihos, J. C. 2009, PASP, 121, 1267
\bibitem[Tal et al.(2009)]{tal09} Tal, T., van Dokkumn, P. G., Jenica, N., \& Bezanson, R. 2009, AJ, 138, 1417
\bibitem[\protect\citeauthoryear{Tapia et al.}{2014}]{Tapia2014} Tapia T., Eliche-Moral M.~C., Querejeta M., Balcells M., C{\'e}sar Gonz{\'a}lez-Garc{\'\i}a A., Prieto M., Aguerri J.~A.~L., et al., 2014, A\&A, 565, A31. doi:10.1051/0004-6361/201321386
\bibitem[Trujillo \& Fliri(2016)]{trujillo16} Trujillo, I., \& Fliri, J. 2016, ApJ, 823, 123
\bibitem[\protect\citeauthoryear{Tonini}{2013}]{Tonini2013} Tonini C., 2013, ApJ, 762, 39. doi:10.1088/0004-637X/762/1/39
\bibitem[\protect\citeauthoryear{Wang et al.}{2012}]{Wang2012} Wang J., Hammer F., Athanassoula E., Puech M., Yang Y., Flores H., 2012, A\&A, 538, A121. doi:10.1051/0004-6361/201117423
\bibitem[Weinberger et al.(2017)]{weinberger17} Weinberger, R., et al. 2017, MNRAS, 465, 3291
\bibitem[Zibetti(2010)]{zibetti10} Zibetti, S. 2010, Astrophysics Source Code Library, ascl:1010.024
\bibitem[Zibetti, Charlot, \& Rix(2009)]{zibetti09} Zibetti, S., Charlot, S., \& Rix, H.--W. 2009, MNRAS, 400, 1181




\end{thebibliography}


\bsp	
\label{lastpage}
\end{document}